\documentclass[trackchanges,twocolumn]{aastex7}

\usepackage{CJK}
\usepackage{amsmath}
\usepackage{xcolor}
\usepackage{bm}
\usepackage{soul}
\usepackage{graphicx}

\graphicspath{figures}

\graphicspath{{./}{figures/}}

\begin{document}
\begin{CJK*}{UTF8}{gbsn}

\title{Properties of current sheets in two-dimensional tearing-mediated incompressible magnetohydrodynamic turbulence}

\correspondingauthor{Chen Shi}
\email{chenshi@auburn.edu}

\author[0000-0002-2582-7085,gname='Chen',sname='Shi']{Chen Shi (时辰)}
\affiliation{Department of Physics, Auburn University, Auburn, AL 36849, USA}
\email{chenshi@auburn.edu}

\author[0000-0002-2381-3106,gname='Marco',sname='Velli']{Marco Velli}
\affiliation{Department of Earth, Planetary, and Space Sciences, University of California, Los Angeles \\
Los Angeles, CA 90095, USA}
\email{mvelli@ucla.edu}

\author[0000-0002-1128-9685,gname='Nikos',sname='Sioulas']{Nikos Sioulas}
\affiliation{Space Sciences Laboratory, University of California, Berkeley\\
Berkeley, CA 94720}
\email{nsioulas@g.ucla.edu}

\author[0000-0002-9968-067X,gname='Zijin',sname='Zhang']{Zijin Zhang}
\affiliation{Department of Earth, Planetary, and Space Sciences, University of California, Los Angeles \\
Los Angeles, CA 90095, USA}
\email{zijin@ucla.edu}

\begin{abstract}
It is well known that the nonlinear evolution of magnetohydrodynamic (MHD) turbulence generates current sheets. In the solar wind turbulence, current sheets are frequently observed and they are believed to be an important pathway for the turbulence energy to dissipate and heat the plasma. 
In this study, we perform a comprehensive analysis of current sheets in a high-resolution two-dimensional simulation of balanced, incompressible MHD turbulence. 
The simulation parameters are selected such that tearing mode instability is triggered and plasmoids are generated throughout the simulation domain. 
We develop an automated method to identify current sheets and accurately quantify their key parameters including thickness ($a$), length ($L$), and Lundquist number ($S$).
Before the triggering of tearing instability, the current sheet lengths are mostly comparable to the energy injection scale. After the tearing mode onsets, smaller current sheets with lower Lundquist numbers are generated.
While power-law scaling relations between $L$ and $a$ and between $a/L$ and $S$ are observed, no clear correlation is found between the upstream magnetic field strength and thickness $a$.
Finally, although the turbulence energy shows anisotropy between the directions parallel and perpendicular to the local magnetic field increment, we do not observe a direct correspondence between the shape of the current sheets and that of the turbulence ``eddies.'' 
These results suggest that one needs to be cautious when applying the scale-dependent dynamic alignment model to the analysis of current sheets in MHD turbulence.

\end{abstract}



\section{Introduction}
It has long been observed that the solar wind is highly turbulent, containing fluctuations on a vast span of scales \citep{bruno2013solar}. 
The solar wind turbulence is believed to be one of the major contributors to the heating and acceleration of solar wind and thus has been extensively investigated for decades \citep[e.g.][]{marsch1990spectral,goldstein1995properties,grappin1996waves,matthaeus2005spatial,chen2012three,alexandrova2013solar,shi2021alfvenic,sioulas2023magnetic}.

One of the key questions regarding solar wind turbulence is how its energy is dissipated.
The widely-accepted picture is that once the energy cascade reaches the end of the inertial range, various processes, such as ion-cyclotron resonance and Landau damping, can take effect and transfer energy from the field to the particles \citep[e.g.][]{cerri2017plasma,he2019direct,martinovic2020enhancement,bowen2022situ}.
In addition to these processes that are important on kinetic scales, intermittency naturally arises in turbulence systems, including neutral fluid \citep[e.g.][]{frisch1980fully} and plasma \citep[e.g.][]{sorriso2001intermittency}. 
Intermittency refers to the sparse, irregular and strong fluctuations on top of the large-scale and relatively smooth fluctuations, and it breaks down the self-similarity of the turbulence, leading to scale-dependent non-Gaussianity of the fluctuations \citep[e.g.][]{papini2020multidimensional}.
In MHD turbulence, intermittency mainly appears in the form of current sheets and vortices \citep{wan2016intermittency,servidio2010statistics}, and it is well-known that current sheets are usually accompanied by strong dissipation via magnetic reconnection \citep{osman2014magnetic}.
Therefore, intermittency may play a significant role in dissipation of MHD turbulence. 
In the solar wind, intermittency has been extensively observed \citep{bruno2001identifying,matthaeus2015intermittency,pei2016influence,wu2023intermittency}, 
and evidence shows that intermittency is co-located with regions of higher proton and electron temperatures in the solar wind \citep{osman2012intermittency,sioulas2022statistical,sioulas2022preferential}.

The question then comes to how do we understand the generation of intermittency in MHD turbulence.
In incompressible-MHD, phenomenological models were developed based on the idea of \textit{scale-dependent dynamic alignment} (SDDA) \citep{boldyrev2005spectrum,podesta2009scale,chandran2015intermittency,mallet2016measures}.
In this scenario, the small-scale turbulence ``eddies'' are assumed to be anisotropic three-dimensionally.
Let us write the eddy size along the background magnetic field $\bm{B_0}$ as $l$, the eddy size along the polarization direction of the fluctuation $\bm{\delta b}$ in the plane perpendicular to $\bm{B_0}$ as $\xi$, and the eddy size perpendicular to both $\bm{B_0}$ and $\bm{\delta b}$ as $\lambda$.
In the perpendicular plane, it was assumed that the eddies are extended along the polarization direction, i.e. $\xi > \lambda$.
A set of scaling relations between these lengths were derived, which were then coupled with the tearing instability theory with the assumption that the ratio $\xi / \lambda$ corresponds to the aspect ratio of the small-scale current sheets \citep{boldyrev2006spectrum,boldyrev2017magnetohydrodynamic,loureiro2017role,mallet2017disruption}.
These authors propose that, as the scale decreases, the aspect ratio may reach a certain critical value, after which the turbulence may be dominated by the recursive collapse of current sheets, i.e. the so-called ``reconnection-mediated'' regime, which is possibly associated with a steepened power spectrum as reported by \citet{dong2018role,dong2022reconnection}.


Although both observations \citep[e.g.][]{podesta2009scale,sioulas2024scale} and numerical simulations \citep[e.g.][]{mason2006dynamic,walker2018influence} have shown evidence of SDDA in plasma turbulence, there is a potential caveat in applying the SDDA theory to the reconnection in turbulence. That is, the turbulence-generated current sheets are intermittent and thus are sparse in space instead of being space-filling like the turbulent eddies.
Hence, whether the anisotropic eddies directly correspond to the shape of intermittent current sheets is questionable.
\citet{dong2018role} conducted high-resolution 2D compressible MHD simulations of balanced turbulence, and found that the dynamic alignment angle in their simulations is too large to explain the aspect ratio of tearing-unstable current sheets.
In this work, to better understand the evolution and properties of current sheets in MHD turbulence and the role of SDDA in this process,
we conduct and analyze a high-resolution 2D incompressible MHD simulation of balanced turbulence with similar setup as \citep{dong2018role}.
The main objective is to compare the scale-dependent dynamic alignment with the properties of intermittent current sheets.
We emphasize that, the classic SDDA theories are not strictly applicable to 2D turbulence due to the lack of propagation effect in 2D ($l \rightarrow +\infty$). 
Nonetheless, we are still able to compute the aspect ratio of the 2D eddies $\xi /\lambda$ and verify whether they are related to the aspect ratio of the current sheets.

The paper is organized as follows: In Section \ref{sec:sim_setup}, we introduce the numerical methods and describe the simulation setup. In Section \ref{sec:results}, we present the results, including the global evolution of turbulence properties (Section \ref{sec:results_evolution_turbulence}), the method to identify and analyze the intermittent current sheets (Section \ref{sec:results_ident_CS}), a case study of an evolving current sheet (Section \ref{sec:results_case}), and the statistics of the current sheet properties (Section \ref{sec:results_statistics}). In Section \ref{sec:discussion}, we discuss the results with a focus on comparing SDDA with current sheet properties. In Section \ref{sec:summary}, we summarize this work.

\section{Simulation setup}\label{sec:sim_setup}
We use the UCLA-Pseudo-Spectral (\texttt{LAPS}) code to conduct the simulation.
\texttt{LAPS} is a Fourier-transform based pseudo-spectral MHD code \citep{shi2024laps}. It has compressible and incompressible versions implemented in both 2D and 3D.
Here, we use the incompressible version which allows us to compare the results with most previous theoretical models that were developed based on the assumption of incompressibility.
We carry out the simulation in 2D so that high resolution can be adopted, which is essential to the development of tearing instability.

The simulation domain (in $x-y$ plane) has the size $L_x = L_y = 1.0$ and numbers of grid points $N_x = N_y = 8192$.
The density is uniform $\rho \equiv 1$ and we set an out-of-plane magnetic field $B_z \equiv 1$. We note that in incompressible 2D MHD, $B_z$ does not affect the evolution and it only serves as a normalization unit in this case. 
We initialize the simulation by adding Alfv\'en-type fluctuations on all wave modes satisfying $ k_{min}\leq \left|k \right| \leq k_{max}$, where $k_{min} = 8$ and $k_{max} = 16$, thus the initial power spectrum is isotropic. 
Here $k=1/\lambda$ with $\lambda$ being the wavelength. 
The amplitude of each individual mode is the same.

\begin{figure}[htb!]
    \includegraphics[width=\linewidth]{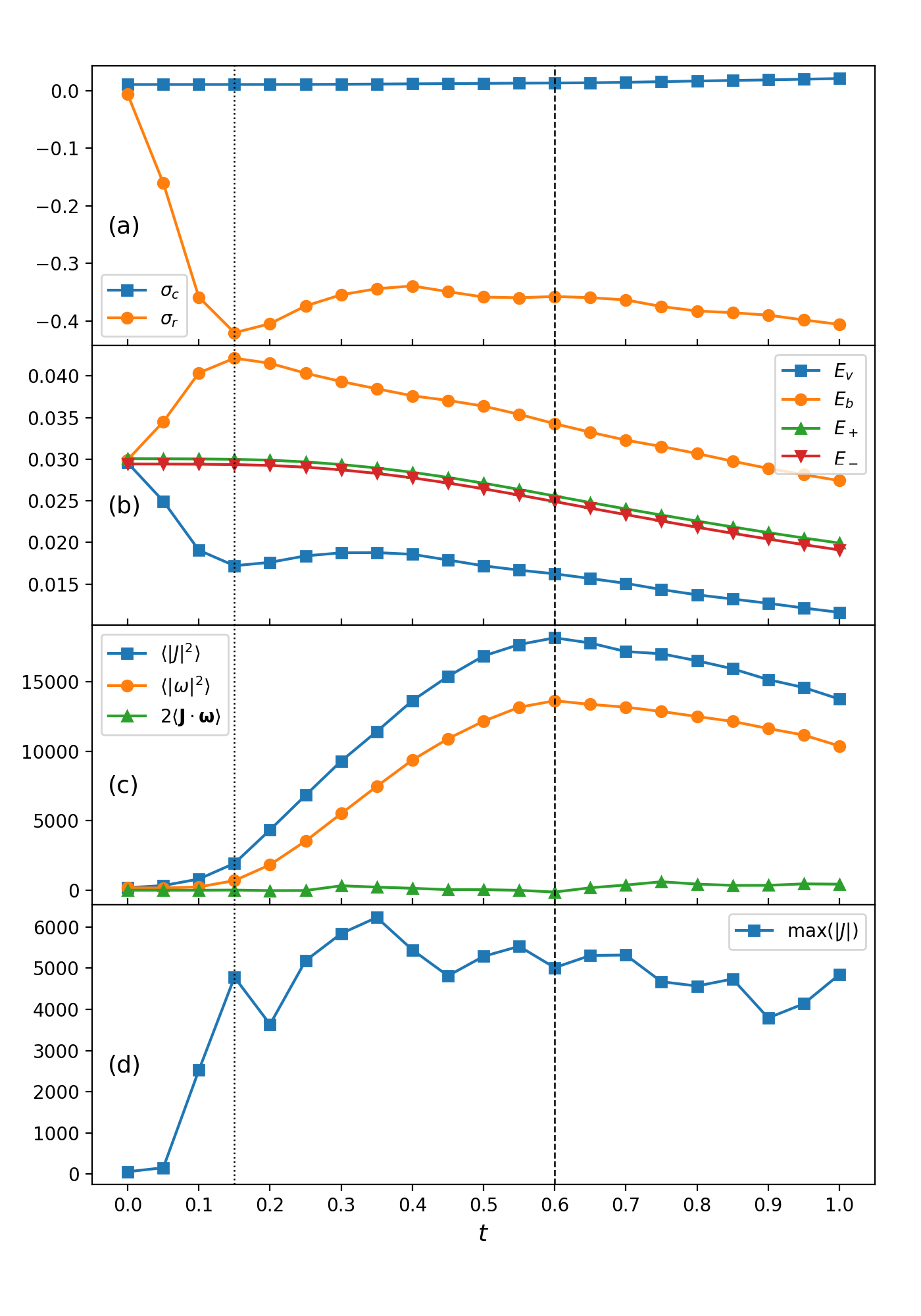}
    \caption{Time evolution of (a) $\sigma_c$ (blue) and $\sigma_r$ (orange); (b) kinetic energy (blue), magnetic energy (orange), and energies of $\bm{z^+}$ (green) and $\bm{z^-}$ (red); (c) averaged $J^2 $ (blue), $\omega^2$ (orange), and $2 \bm{J} \cdot \bm{\omega}$ (green); (d) Maximum $|J|$ in the simulation domain.}
    \label{fig:evolution_global_quantities}
\end{figure}

\begin{figure*}
    \centering
    \includegraphics[width=\hsize]{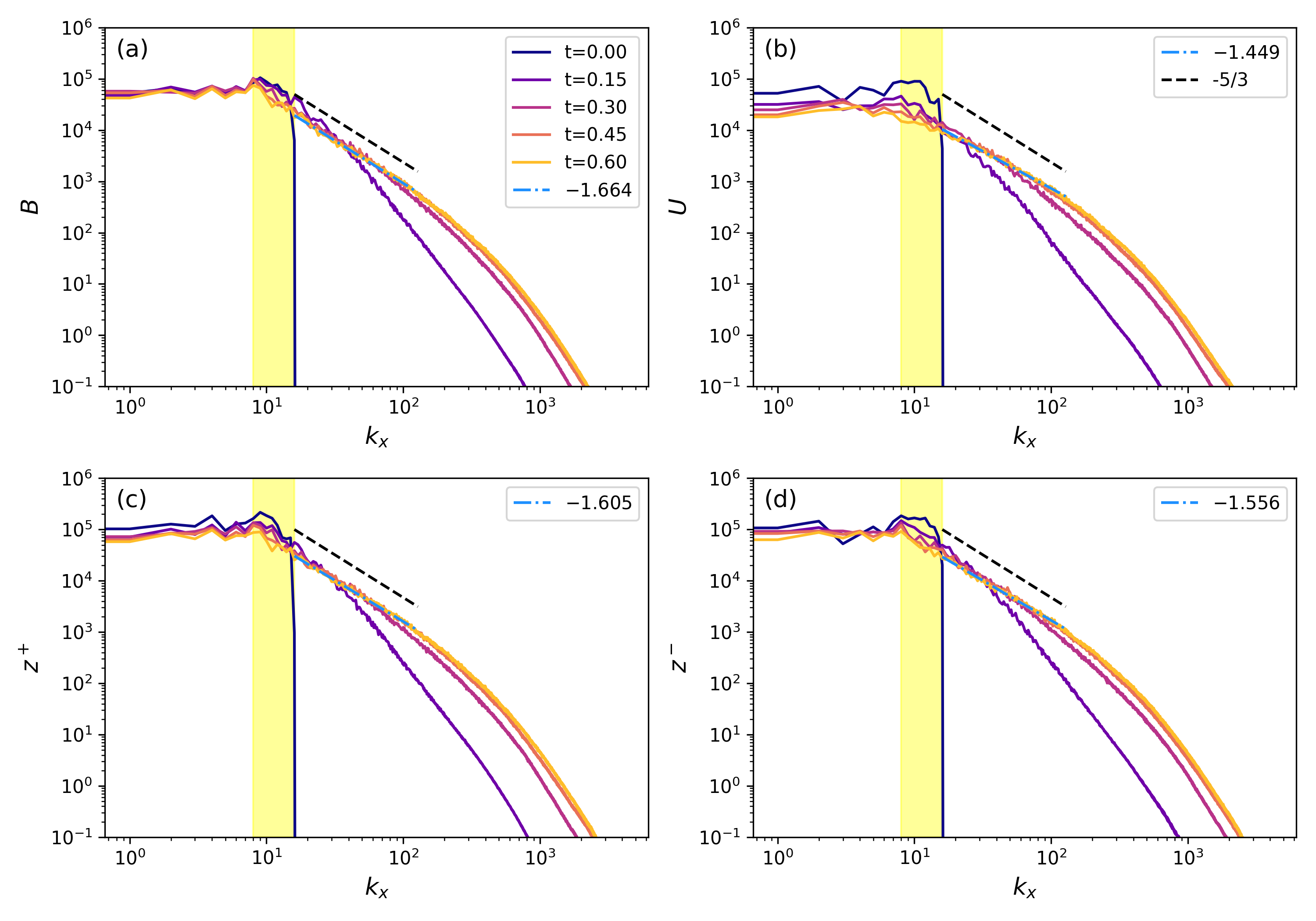}
    \caption{Power spectra of (a) magnetic field, (b) velocity, (c) $\bm{z^+}$, and (d) $\bm{z^-}$ calculated along $x$. In each panel, different solid curves correspond to spectra at different time moments. Blue dotted-dashed lines are linear-fitting of the spectra at $t=0.6$ using $16\leq k \leq 128$. The black dashed lines show $\propto k^{-5/3}$ for reference. The yellow shade in each panel marks $8 \leq k \leq 16$ that correspond to the initial fluctuations.}
    \label{fig:power_spectra}
\end{figure*}

For each $\bm{k}$, the perturbation satisfies $\bm{b_{k}} \perp \bm{k}$, and we add the two Alfv\'en modes, i.e. $\bm{v_k } = \pm \bm{b_k}$, with the same amplitude but random phases.
Consequently, the normalized cross helicity 
\begin{displaymath}
    \sigma_c = \frac{E_+ - E_-}{E_+ + E_-}
\end{displaymath}
and normalized residual energy 
\begin{displaymath}
    \sigma_r = \frac{E_v - E_b}{E_v + E_b}
\end{displaymath} 
are both approximately zero.
Here $E_v = \int v^2 $, $E_b = \int b^2$, and $E_\pm = \int {z^\pm}^2$ measure the total energies in velocity, magnetic field, and the two Els\"asser variables $\bm{z^\pm} = \bm{v} \mp \bm{b}$.
The root-mean-square amplitude of the initial fluctuation is $RMS(\bm{b}) = 0.173$.
Explicit resistivity $\eta$ and viscosity $\nu$ are implemented with $\eta = \nu = 5\times10^{-7}$.
Thus the (magnetic) Reynolds number is estimated to be $S_0 = \lambda \delta b/\eta \approx 4 \times 10^4$ using $\lambda = 1/k_{min}$.
Theoretically, the maximum Reynolds number that a turbulence simulation can resolve is $S_{max} \sim (L/d)^{4/3}$ where $L$ and $d$ are the outer scale and dissipation scale respectively.
Take the outer scale $L=1/k_{min}=1/8$ and the dissipation scale $d=1/N_x=1/8192$, we get $S_{max}\approx1\times 10^4 $, slightly smaller than but on the same order of $S_0$.
To verify that the selected $\eta$ and $\nu$ are within a reasonable range, we conducted tests with different amplitudes of dissipation, including $\eta = \nu = \{1\times 10^{-7},\, 5\times 10^{-7}  ,\,  1\times 10^{-6}$\}, and the values $\eta = \nu = 5\times10^{-7}$ are chosen because we observe that the dissipation is strong enough to effectively suppress the numerical error induced by Gibbs phenomenon while not too strong so that a significant number of tearing-unstable current sheets are generated.
For a comprehensive discussion of how the numerical resistivity and viscosity affect the turbulence evolution, such as the number of X-points, please refer to \citep{wan2013generation}.
The eddy-turnover time $\tau_{nl} = \lambda / z_\lambda \approx 0.7$. Thus, we run the simulation to $t=1.0$, roughly 1.4 eddy-turnover times.
A compressible run with thermal pressure $P=0.5$, i.e. $\beta = 1$, was also conducted. The evolution, e.g. the energy dissipation rate, does not change significantly from the incompressible run (the compressible run has slightly lower dissipation rate). However, a quantitative comparison between the compressible and incompressible runs will be necessary to fully understand how compressibility modifies the evolution.

\begin{figure*}
    \centering
    \includegraphics[width=\hsize]{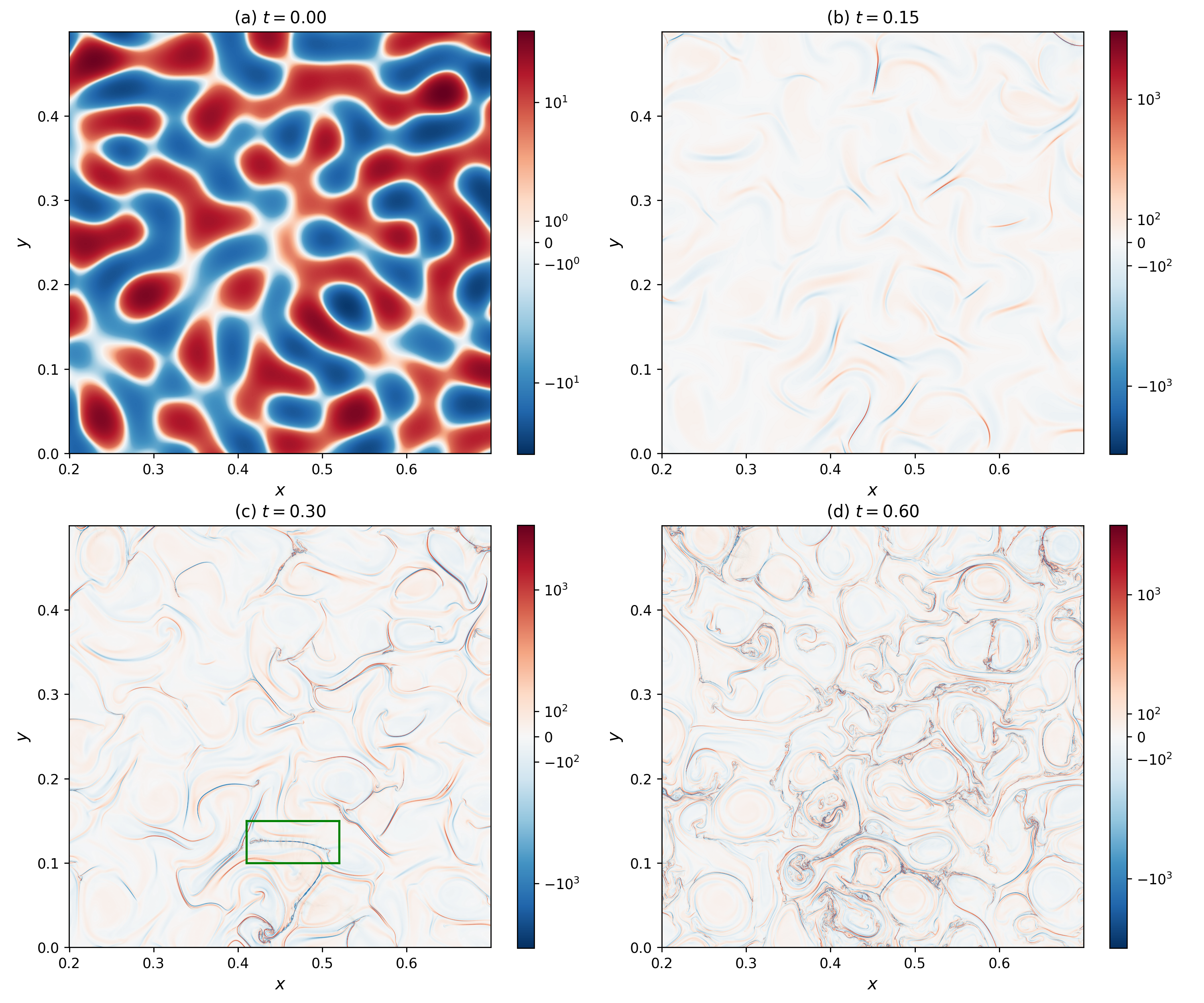}
    \caption{Evolution of $J_z$ in the subdomain $x \in [0.2,0.7]$, $y\in [0,0.5]$. In Panel~(c), the green box marks the current sheet analyzed in detail in Section~\ref{sec:results_case} and Figure~\ref{fig:evolution_single_CS}.}
    \label{fig:evolution_Jz_2D}
\end{figure*}

\section{Results}\label{sec:results}

\subsection{Evolution of turbulence properties}\label{sec:results_evolution_turbulence}

In Figure~\ref{fig:evolution_global_quantities}, we show the time evolution of various quantities. Panel~(a) shows that $\sigma_c$ (blue) remains roughly zero throughout the simulation, while $\sigma_r$ (orange) decreases rapidly at the early stage ($t\leq 0.15$), then increases a little bit and finally starts to decay slowly.
The initial rapid decrease of $\sigma_r$ was observed in 3D simulations with similar setup \citep{shi2025evolution} as well as 2D hybrid simulations \citep[e.g.][]{franci2015high}.
In Panel~(b), we show the evolution of $E_v$ (blue), $E_b$ (orange), $E_+$ (green), and $E_-$ (red) respectively. 
$E_+$ and $E_-$ decay slowly due to the numerical dissipation, and so for $E_v$ and $E_b$ after $t \approx 0.3$.
The early stage ($t \leq 0.15$) is featured by a rapid increase in $E_b$ and decrease in $E_v$, leading to the decrease in $\sigma_r$.
We mark the peak of $E_b$ (and the valley of $E_v$) with the vertical dotted line.
Clearly, there is a fast transfer of kinetic energy to magnetic energy at the early stage, much earlier than one eddy-turnover time.
As will be shown in Figure~\ref{fig:evolution_Jz_2D}, thin current sheets are formed concurrently.
After $t=0.15$, reconnection becomes significant in these current sheets, transferring magnetic energy back to the kinetic energy, which probably explains the slight increase in $E_v$ after $t=0.15$.
As will be shown in Figure~\ref{fig:sigma_r_2D}, the negative residual energy is concentrated near the eddy boundaries, in regions surrounding the thin current sheets, rather than inside the current sheets where plasma jets are generated and thus featured by positive residual energy.

Panel~(c) shows evolution of energy dissipation rates, i.e. the magnetic energy dissipation rate $\langle |\bm{J}|^2 \rangle$ (blue) with $\bm{J} = \nabla\times \bm{b}$ being the current density, the kinetic energy dissipation rate $\langle |\bm{\omega}|^2\rangle$ (orange) with $\bm{\omega} = \nabla\times \bm{v}$ being the vorticity, and the cross term $2 \langle \bm{J} \cdot \bm{\omega} \rangle$ (green).
In this simulation, $\bm{J} = J_z \hat{e}_z$ and $\bm{\omega} = \omega_z \hat{e}_z$.
We note that the dissipation rates for $\bm{z^\pm}$ are $\langle |\bm{J}|^2 + |\bm{\omega}|^2 \mp 2 \bm{J} \cdot \bm{\omega} \rangle$ respectively.
Here $\langle \rangle$ represents average over the whole simulation domain.
The vertical dashed line marks $t=0.6$, corresponding to the peaks of $\langle \bm{J}^2 \rangle$ and $\langle \bm{\omega}^2 \rangle$. This is the time moment when the turbulence is fully developed. After this moment, the turbulence starts to decay. 
Note that 0.6 is close to one eddy turnover time $\tau_{nl} \approx 0.7$.
In panel~(d), we plot the evolution of max$(|J|)$ in the simulation domain, which is often used as an indicator of the onset of magnetic reconnection \citep{franci2017magnetic,papini2019can}. One can see that max$(|J|)$ shows a peak at $t=0.15$, implying that reconnection onsets around this time moment, consistent with other evidence such as $E_b$.

\begin{figure}
    \centering
    \includegraphics[width=\linewidth]{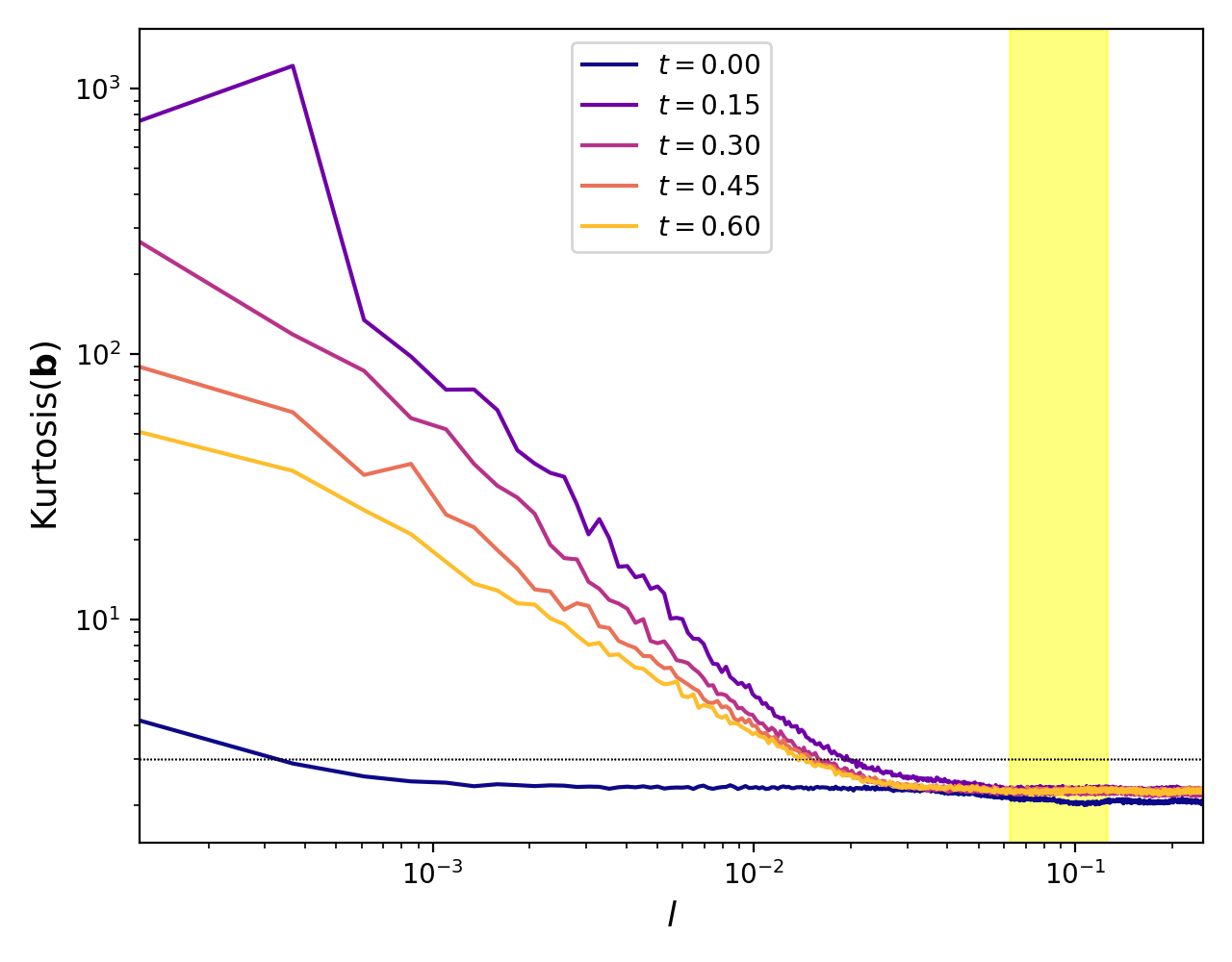}
    \caption{Scale-dependent Kurtosis of magnetic field at different time moments. The yellow shade marks the wavelength-range of the initial fluctuations.}
    \label{fig:kurtosis_b}
\end{figure}

\begin{figure*}
    \centering
    \includegraphics[width=\linewidth]{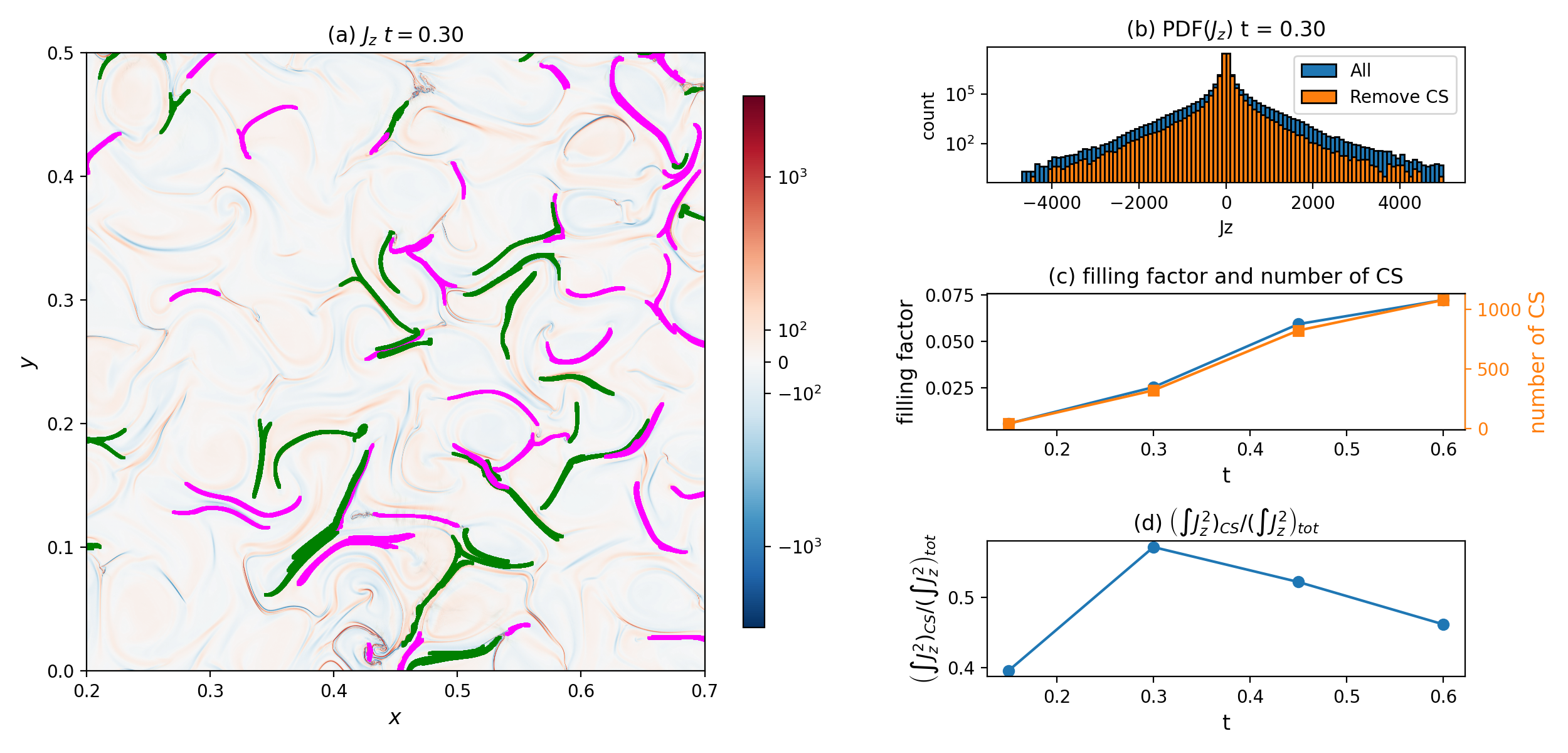}
    \caption{(a) Same as Panel~(c) of Figure~\ref{fig:evolution_Jz_2D} with identified current sheets marked by magenta ($J_z>0$) and green ($J_z<0$). (b) Probability distribution function (PDF) of $J_z$ at $t=0.3$. Blue bars are all the data points and orange bars are data points with identified current sheets removed. (c) Time evolution of filling factor (blue), i.e. area of current sheets divided by the total area of the domain, and number of current sheets (orange). (d) $ \int J_z^2 $ inside the current sheets over $\int J_z^2 $ throughout the simulation domain.}
    \label{fig:Jz2D_filling_factor_etc}
\end{figure*}

In Figure~\ref{fig:power_spectra}, we show the power spectra for different quantities at different time moments. We fit the spectra at $t=0.6$, when the turbulence is fully developed and the spectra are stably established over $ 16 \leq k \leq 128$, which is shown by the blue dotted-dashed lines. 
The fitted slopes are written in the legends. The magnetic field spectral slope is close to $-5/3$ and the velocity spectrum is shallower. This has been confirmed by a large number of numerical simulations \citep[e.g.][]{muller2005spectral,boldyrev2011spectral,papini2019can,shi2025evolution} and observations \citep[e.g.][]{podesta2007spectral,chen2013residual,shi2021alfvenic}.
The slopes for $\bm{z^\pm}$ are similar and take values between $-3/2$ and $-5/3$.
Figure~\ref{fig:power_spectra} confirms that the formation of thin current sheets ($t=0.15$, Figure~\ref{fig:evolution_global_quantities}) is well before the turbulence is fully developed, since the power spectra are still evolving at $t=0.15$.

In Figure~\ref{fig:evolution_Jz_2D}, we show four snapshots of $J_z$ in the subdomain $x \in [0.2,0.7], \, y\in[0,0.5]$.
Panel~(a) shows the initial condition, where $J_z$ is smooth (note that the color bar is in log-scale).
Panel~(b) shows $t=0.15$, corresponding to the vertical dotted line in Figure~\ref{fig:evolution_global_quantities}. At this moment, a number of thin current sheets are generated, but fast reconnection has not onset yet.
Panel~(c) shows $t=0.3$, when more current sheets are generated and the current sheets become thinner than $t=0.15$.
Tearing instability has been triggered in some of the current sheets, indicated by the generation of plasmoids. One tearing-unstable current sheet is marked by the green box and this current sheet will be analyzed in detail in Section \ref{sec:results_case}.
Panel~(d) shows $t=0.6$, corresponding to the vertical dashed line in Figure~\ref{fig:evolution_global_quantities}, when the turbulence is fully developed. Much more plasmoids are generated at this time compared with $t=0.3$.
Figure~\ref{fig:evolution_Jz_2D} shows that the generated current sheets are located near the boundaries of eddies whose sizes are close to to the wavelengths of the initial fluctuations.

In Figure~\ref{fig:kurtosis_b}, we show the scale dependent Kurtosis of the magnetic field, defined by
\begin{equation}
    \mathrm{Kurtosis}(\bm{b},l) = \frac{ \langle  \left| \delta \bm{b}(l)  \right|^4 \rangle}{\langle \left| \delta \bm{b}(l) \right|^2 \rangle^2},
\end{equation}
at different time moments. 
Here $\delta\bm{b}=\bm{b}(\bm{x}+\bm{l}) - \bm{b}(\bm{l})$.
In general, the smaller scales correspond to larger values of Kurtosis , which is a typical feature of intermittent systems \citep{papini2020multidimensional}. 
Interestingly, it is observed that the Kurtosis reaches maximum at $t=0.15$ and then gradually decay.
Combined with Figure~\ref{fig:evolution_global_quantities}, it indicates that the increase of Kurtosis is likely associated with the current sheet formation process, and the later decrease of the Kurtosis is potentially due to reconnection disrupting the current sheets.

\subsection{Identification of current sheets}\label{sec:results_ident_CS}

\begin{figure*}[htb!]
    \centering
    \includegraphics[width=\linewidth]{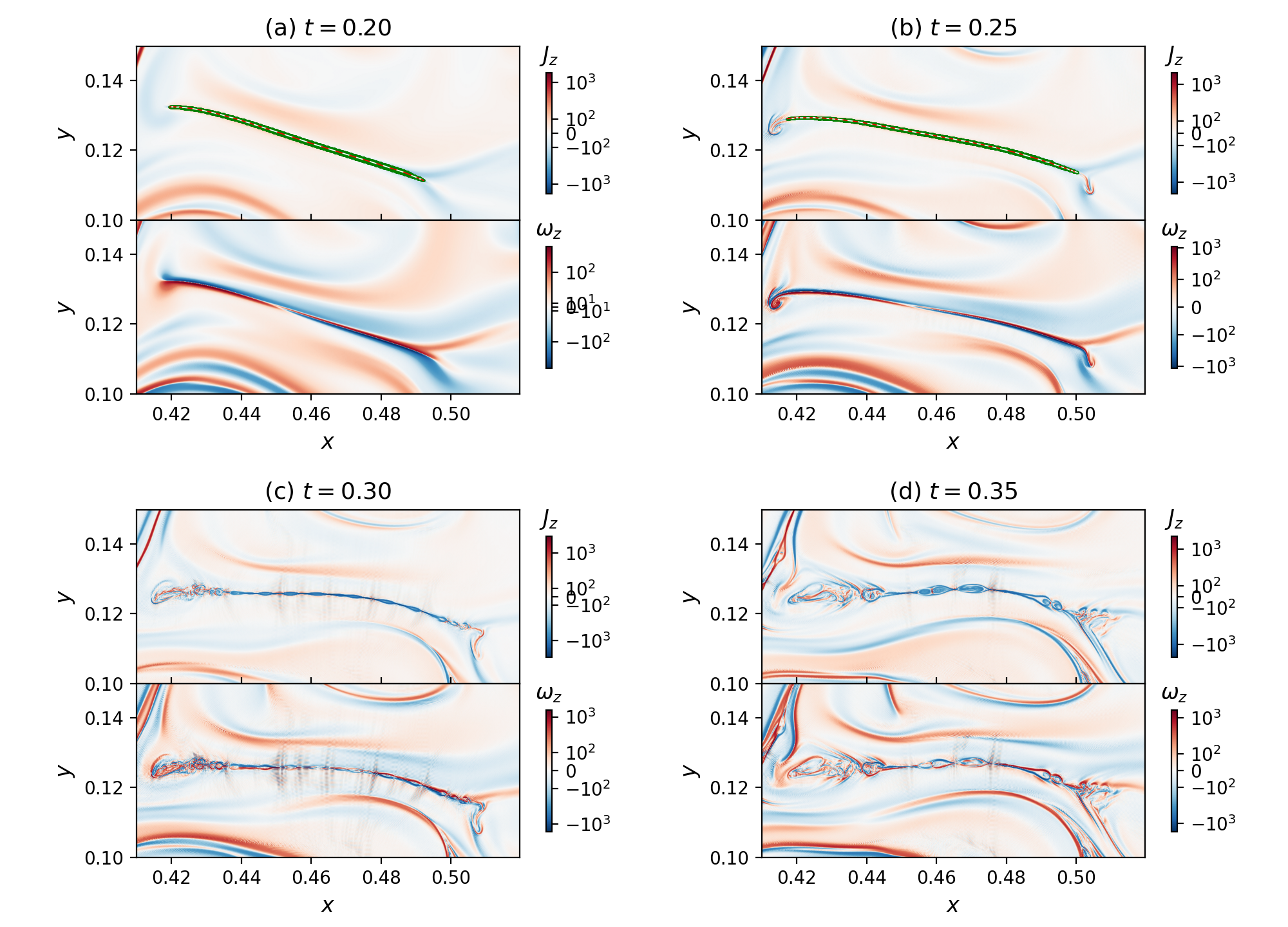}
    \caption{Time evolution of a single current sheet within the domain marked by the green box in Figure~\ref{fig:evolution_Jz_2D}. In each panel, top plot shows $J_z$ and bottom plot shows $\omega_z$.
    In panels~(a) \& (b), the green shade is the automatically identified current sheet region using the algorithm described in Section~\ref{sec:results_ident_CS}.
    The white and red lines correspond to the segments of the current sheet given by the recursive PCA algorithm described in Section~\ref{sec:results_case}.}
    \label{fig:evolution_single_CS}
\end{figure*}

\begin{figure}
    \centering
    \includegraphics[width=\linewidth]{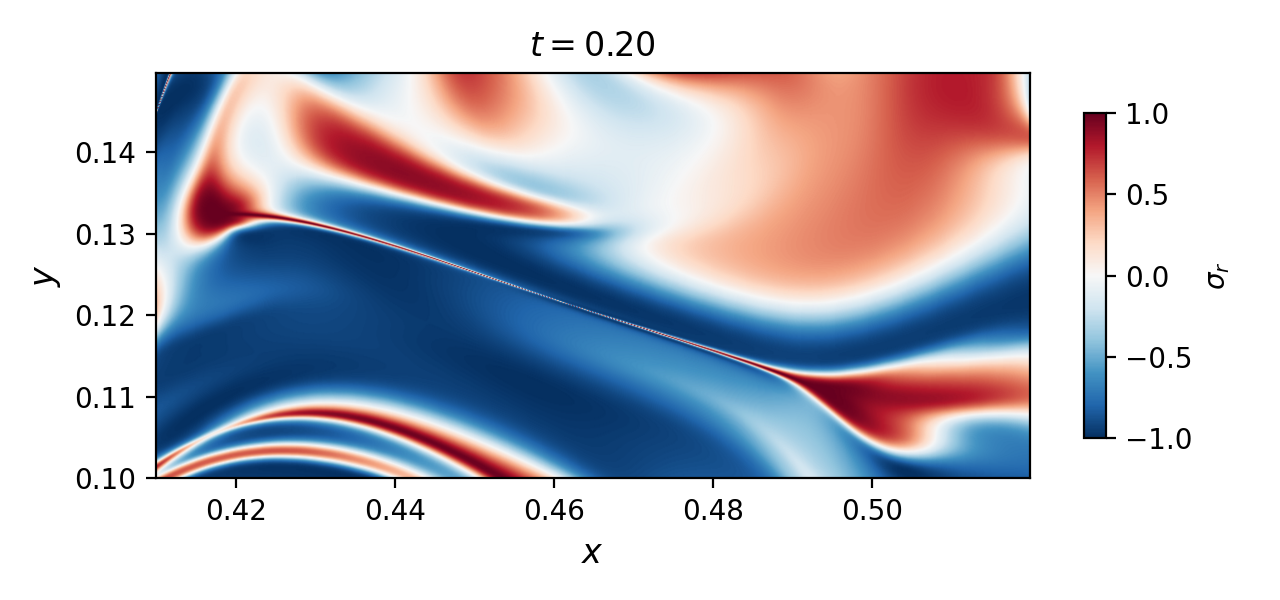}
    \caption{$\sigma_r$ at $t=0.20$ around the current sheet shown in Figure~\ref{fig:evolution_single_CS}.}
    \label{fig:sigma_r_2D}
\end{figure}

To quantify the properties of the current sheets, it is necessary to accurately identify the current sheets.
We develop an algorithm similar to that described in \citep{zhdankin2013statistical}, and we briefly describe our algorithm here.
We start from finding the grid point corresponding to $\max(\left| J_z \right|)$ throughout the whole domain, excluding regions already marked as current sheets.
Then we run a recursive, depth-first search algorithm to check the four neighboring grid points. 
The algorithm returns when all the neighboring points have $\left| J_z \right| \leq 0.1 \max(\left| J_z \right|)$, and we mark the continuous set of data points as one single current sheet.
The threshold influences the area of the identified current sheet: a lower value leads to a larger current sheet area. There is no standard for this threshold, but 0.1 is good enough to identify coherent current sheets, as will be shown by Figures~\ref{fig:Jz2D_filling_factor_etc} and \ref{fig:evolution_single_CS}. 
This current-sheet search is repeated until all the current sheets with peak-$\left| J_z \right|$ larger than $0.1 \max_{global}(\left| J_z \right|)$ are identified, where $\max_{global}(\left| J_z \right|)$ is the maximum $\left| J_z \right|$ in the whole simulation domain.
Obviously, the threshold affects the total number of current sheets identified. We select 0.1 such that a sufficient number of current sheets are identified for solid statistics, while it does not take too long to run the algorithm.
In Figure~\ref{fig:Jz2D_filling_factor_etc}(a), we show the same plot as Figure~\ref{fig:evolution_Jz_2D}(c) and mark the identified current sheets in magenta ($J_z>0$) and green ($J_z<0$). 
It shows that the algorithm works well, and most of the identified current sheets exist between the interacting large-scale ``eddies'', i.e. near the eddy boundaries, as pointed out by \citet{papini2019can} in their Hall-MHD and hybrid Particle-in-Cell simulations.
However, we note that, as the turbulence develops, there are many regions with strong current density but are ``broken'' instead of being coherent structures.
Thus, we further refine the current sheet lists by discarding regions with less than 20 data points.

In Panel~(b) of Figure~\ref{fig:Jz2D_filling_factor_etc}, we show the probability distribution function (PDF) of $J_z$ at $t=0.3$. Here blue bars are all the data points in the whole simulation domain and orange bars are points after removing the identified current sheets.
The PDF without current sheets is naturally thinner than that with current sheets included.
Panel~(c) displays the filling factor (blue) and the number (orange) of current sheets. The filling factor is defined as the total area of the identified current sheets divided by the area of the simulation domain.
Panel~(d) shows the normalized contribution of the identified current sheets to the total magnetic energy dissipation, i.e. ratio between the integrated $J_z^2$ in the current sheets and the integrated $J_z^2$ throughout the whole simulation domain.
Panels~(c) and~(d) show that, the current sheets, with a total area less than 10\% of the simulation domain, contribute roughly 50\% of the dissipation of magnetic energy.
Similar conclusion was made by applying a space-filtering technique to derive the spectral energy flux in a 2D Hall-MHD simulation \citep{camporeale2018coherent}.

\subsection{Case study of a current sheet}\label{sec:results_case}
We aim to accurately quantify the geometry of these current sheets, including their lengths and thicknesses.
Therefore, a principal component analysis (PCA) based algorithm is developed, with the aid of PCA functions implemented in the Python package Scikit-Learn \citep{scikit-learn}.

For a set of spatial coordinates $(x_i,y_i)$ ($i=0,\cdots N-1$), PCA determines two orthogonal unit vectors, along which the set of coordinates has maximum and minimum variances respectively.
We refer to the axes along the two unit vectors as major (maximum variance) and minor (minimum variance) axis.
For a straight current sheet, the major and minor axes roughly correspond to its length and thickness.
However, as can be seen from Figure~\ref{fig:Jz2D_filling_factor_etc}(a), most of the current sheets are curved, hence increasing the error in quantifying the current sheet length.
To reduce this error, we apply PCA recursively for each current sheet. That is, after applying PCA, we check the ratio between the maximum variance and minimum variance. 
If this ratio is larger than 3, it means the data points are still quite spread in a long and thin region, and we further divide the dataset into two subsets, i.e. left and right of the center of mass $(\langle x_i \rangle, \langle y_i \rangle)$ along the major axis, and apply PCA to the two subsets separately until the ratio between maximum and minimum variances gets smaller than 3.
With this recursive-PCA algorithm, we are able to break a long and curved current sheet into multiple segments along the major axis.
For each segment, we define the length as the extent of the data points along the major axis, and the length of the current sheet is the sum-up of the lengths of all the segments. 

\begin{figure*}
    \centering
    \includegraphics[width=\linewidth]{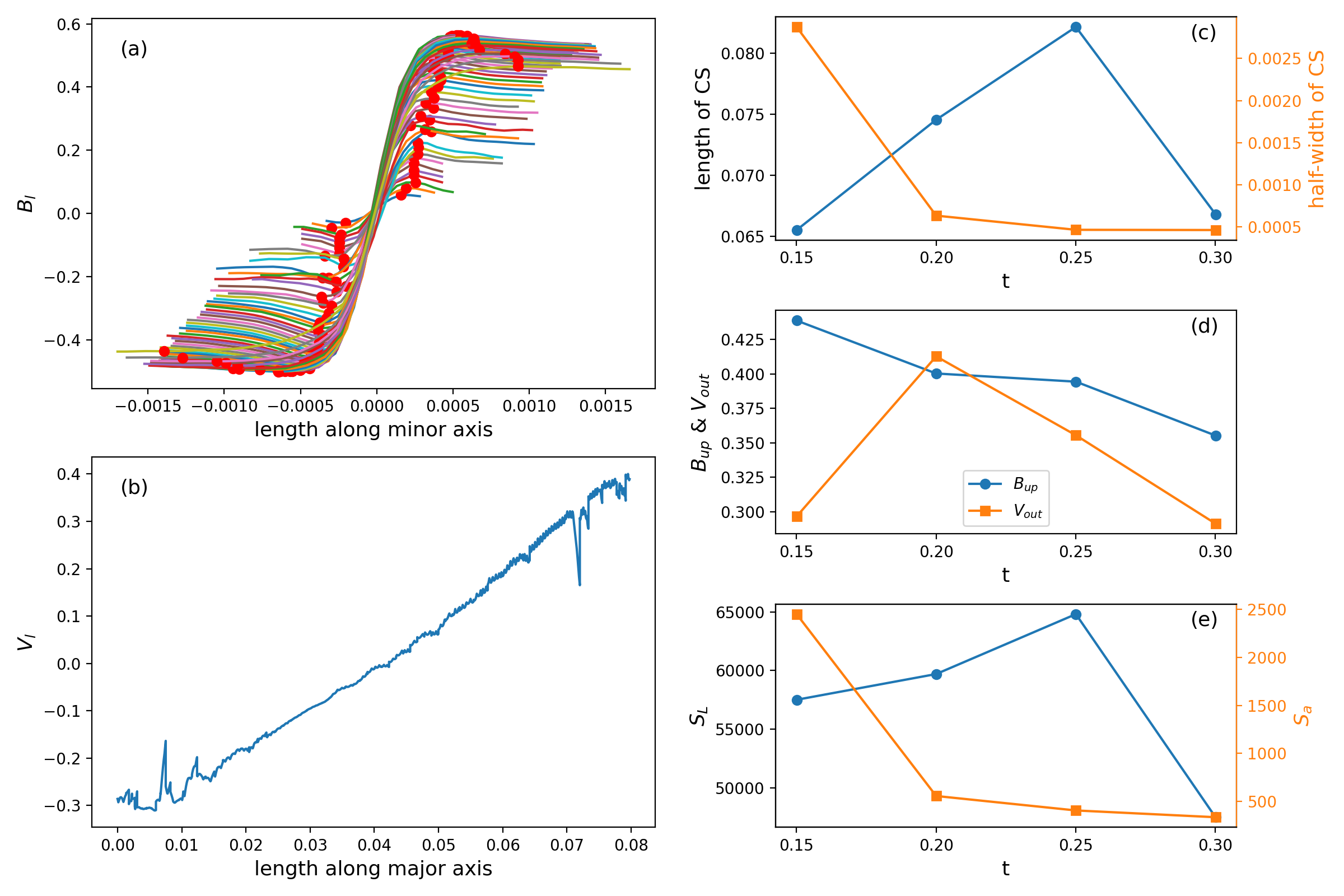}
    \caption{(a) $B_l$ as a function of distance along the minor axis at $t=0.25$. Different curves correspond to different segments (see the text) in the recursive PCA analysis. Red circles mark the maximum and minimum of each $B_l$ profile. (b) $V_l$ as a function of distance along the major axis, at the center of the current sheet, at $t=0.25$. (c) Time evolution of the length (blue circle) and half-width (orange square) of the current sheet. (d) Time evolution of the upstream magnetic field strength $B_{up}$ and the outflow speed $V_{out}$. (e) Time evolution of the Lundquist numbers $S_L = L B_{up}/\eta$ (blue circle) and $S_a = a B_{up}/\eta$ (orange square).}
    \label{fig:single_CS_PCA}
\end{figure*}

In Figure~\ref{fig:evolution_single_CS}, we show evolution of the current sheet marked by the green box in Figure~\ref{fig:evolution_Jz_2D}(c). 
At each time moment, we plot $J_z$ on the top and $\omega_z$ on the bottom.
In panels~(a) \& (b), the identified current sheet region is marked by the green shade.
The short white and red lines correspond to the segments given by the recursive PCA algorithm.
Clearly, from $t=0.2$ to $t=0.25$, the current sheet is stretched and thinned. 
The vorticity shows a quadrupole structure, implying bidirectional jets, which is evidence of ongoing reconnection.
In Figure~\ref{fig:sigma_r_2D}, we show $\sigma_r$ around this current sheet at $t=0.20$. As mentioned in Section~\ref{sec:results_evolution_turbulence}, $\sigma_r$ is negative on the two sides of the current sheet and positive inside the current sheet because of the plasma jets.
At $t=0.25$, signature of shear-driven instability is observed at the tips of the current sheet.
At $t=0.3$, tearing instability is triggered and enters the nonlinear stage, generating multiple magnetic islands. We estimate the average length and half-width of the magnetic islands to be $\lambda \approx 5 \times 10^{-3}$ and $w \approx 5 \times 10^{-4}$ respectively.
Later, at $t=0.35$, the tearing instability continues to develop nonlinearly, featured by coalescing magnetic islands.


In Figure~\ref{fig:single_CS_PCA}, we show the PCA result of the current sheet shown in Figure~\ref{fig:evolution_single_CS}.
In panel~(a), we show $B_l$, i.e. magnetic field projected along the major axis, as a function of the coordinate along minor axis at $t=0.25$. Each curve corresponds to one segment in the PCA analysis. 
One can see that the $B_l$ profile is similar to Harris type but its amplitude decreases far away from the center of the current sheet.
For each profile, we mark the maximum/minimum of $B_l$ with the red circles, and take the distance between the two red circles as the width (thickness) of the current sheet segment.
We then average the widths of all the segments and define this average value as the width of the current sheet.
In panel~(b), we show the variation of $V_l$ along the major axis at the center of the current sheet at $t=0.25$. $V_l$ changes from $-0.3$ to $0.4$ as we move from the left to the right side of the current sheet.
Panel~(c) shows the time evolution of the length $L$ (blue circles) and half-width $a$ (orange squares) of the current sheet. For $t\leq0.25$, the current sheet lengthens and thins until $L \approx 0.08$ and $a \approx 0.005$, after which it enters the nonlinear tearing stage.
Panel~(d) shows the time evolution of the upstream magnetic field $B_{up}$ (blue), defined as $\langle B_{l,max} - B_{l,min}\rangle/2$ where $B_{l,max}$ and $B_{l,min}$ correspond to the red circles and $\langle \rangle$ means average over all segments, and the outflow speed $V_{out}$, defined as $(V_{l,max} - V_{l,min})/2$ along the major axis as shown by panel~(b).
Before the nonlinear tearing stage, the two quantities are similar. As the tearing mode grows, the reconnection is no longer laminar, and $V_{out}$ becomes smaller than $B_{up}$.
Last, in panel~(e), we show the evolution of the estimated Lundquist numbers $S_L = L B_{up} / \eta$ and $S_a = a B_{up} / \eta$. 
In the linear-tearing stage ($t=0.25$), we have $S_L \approx 6.5 \times 10^4$ and $S_a \approx 500$.
Note that at this moment $a/L \approx 5\times10^{-4}/0.08 = 0.00625 \approx S_L^{-0.46}$, close to the Sweet-Parker model which predicts $a/L \sim S_L^{-1/2}$.

In summary, we conclude that this single current sheet undergoes the following stages: (1) Formation by thinning and lengthening accompanied by laminar reconnection. (2) Growth of linear tearing instability when the aspect ratio approaches the Sweet-Parker scaling $a/L \sim S_L^{-1/2}$. (3) Fast dissipation of the current sheet by nonlinear tearing instability. 
We note that, the Lundquist number $S_L= 6.5 \times 10^4$ of this current sheet is not significantly larger than the critical Lundquist number, approximately $O(10^4)$, for the Sweet-Parker type current sheet to be tearing-unstable \citep[e.g.][]{biskamp1986magnetic,loureiro2007instability,shi2018marginal}.
In the nearly-collisionless space plasma with much larger Lundquist numbers, the growth of linear tearing mode may happen earlier, when the aspect ratio $a/L$ approaches $S_L^{-1/3}$ instead of $S_L^{-1/2}$ \citep{pucci2013reconnection,tenerani2015magnetic,tenerani2016ideally,del2016ideal,papini2019fast}.

\subsection{Statistics of current sheets}\label{sec:results_statistics}

\begin{figure*}[htb!]
    \centering
    \includegraphics[width=\hsize]{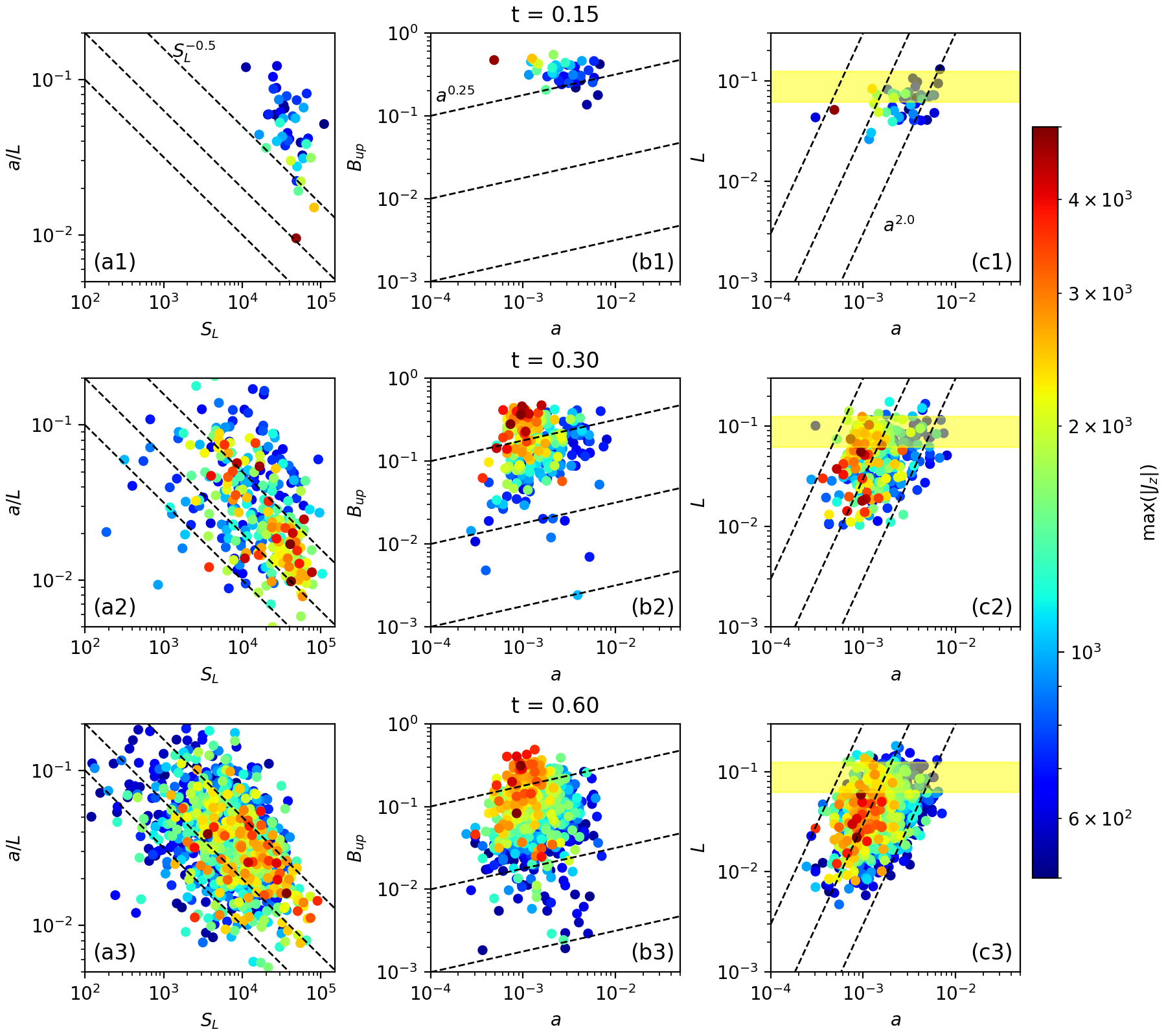}
    \caption{Statistics of the current sheet properties. Top to bottom rows show $t=0.15$, $0.30$, and $0.60$ respectively. Left column (a) shows the aspect ratio $a/L$ as a function of Lundquist number $S_L$. Middle column (b) shows the upstream magnetic field $B_{up}$ as a function of half-width of the current sheet $a$. Right column (c) shows the current sheet length $L$ as a function of its half-width $a$. Each dot corresponds to one single current sheet and is color-coded with the peak current density inside the current sheet. In the right column, the yellow shades mark $1/16 \leq L \leq 1/8$.}
    \label{fig:statistics_current_sheet}
\end{figure*}

We apply the analysis described in Section~\ref{sec:results_case} to all the identified current sheets and present the results in Figure~\ref{fig:statistics_current_sheet}.
Top to bottom rows show time moments $t=0.15$, $0.3$, and $0.6$ respectively.
Left column shows aspect ratio $a/L$ versus Lundquist number $S_L$, middle column shows upstream magnetic field $B_{up}$ versus current sheet thickness $a$, and right column shows current sheet length $L$ versus thickness $a$.
Each dot corresponds to one identified current sheet, and the dot is color-coded by the peak current density in the current sheet.
In the left column, we plot reference lines $a/L \propto S_L^{-1/2}$, i.e. the Sweet-Parker type current sheet model.
We can see that for all the three moments, the data points show a negative correlation between $a/L$ and $S_L$, and they roughly obey $a/L \sim S_L^{-1/2}$.
However, we emphasize that, because the data points are quite scattered, the scaling relation between $a/L$ and $S_L$ cannot be accurately determined. An ideal-tearing scaling may be obtained in a simulation with much higher Lundquist number \citep{papini2019fast}. But this requires further investigation.
In the middle column, we plot reference lines $B_{up} \propto a^{1/4}$, which is the prediction by the SDDA theory \citep{boldyrev2006spectrum}.
One can see that there is no clear correlation between $B_{up}$ and $a$.
In contrast, the right column shows a strong positive correlation between $L$ and $a$.
For reference, we plot $L \propto a^2$, which is the scaling relation given by Sweet-Parker model, i.e. $a/L \sim (L B_{up}/\eta)^{-1/2}$, assuming that the upstream Alfv\'en speed is independent of $a$.  
Similarly, for ideal-tearing scaling relation $a/L \sim (L B_{up}/\eta)^{-1/3}$, we will get $L \propto a^{3/2}$.
Note that, SDDA theory predicts $L \propto a^{3/4}$, much shallower than what is shown here.
We point out that, at $t=0.15$, i.e. before tearing instability is triggered, most of the current sheets have lengths close to the wavelengths of initial fluctuations marked by the yellow shades.
Moreover, panel~(b1) shows that the upstream magnetic field is also quite consistent among all the current sheets, slightly above the initial fluctuation level ($\delta b  \approx 0.17$).
These results indicate that the generation of current sheets is dominated by the largest turbulence eddies in the system \citep{papini2019can,khan2025does}, and their subsequent evolution is likely dominated by tearing instability.

\begin{figure*}[htb!]
    \centering
    \includegraphics[width=\hsize]{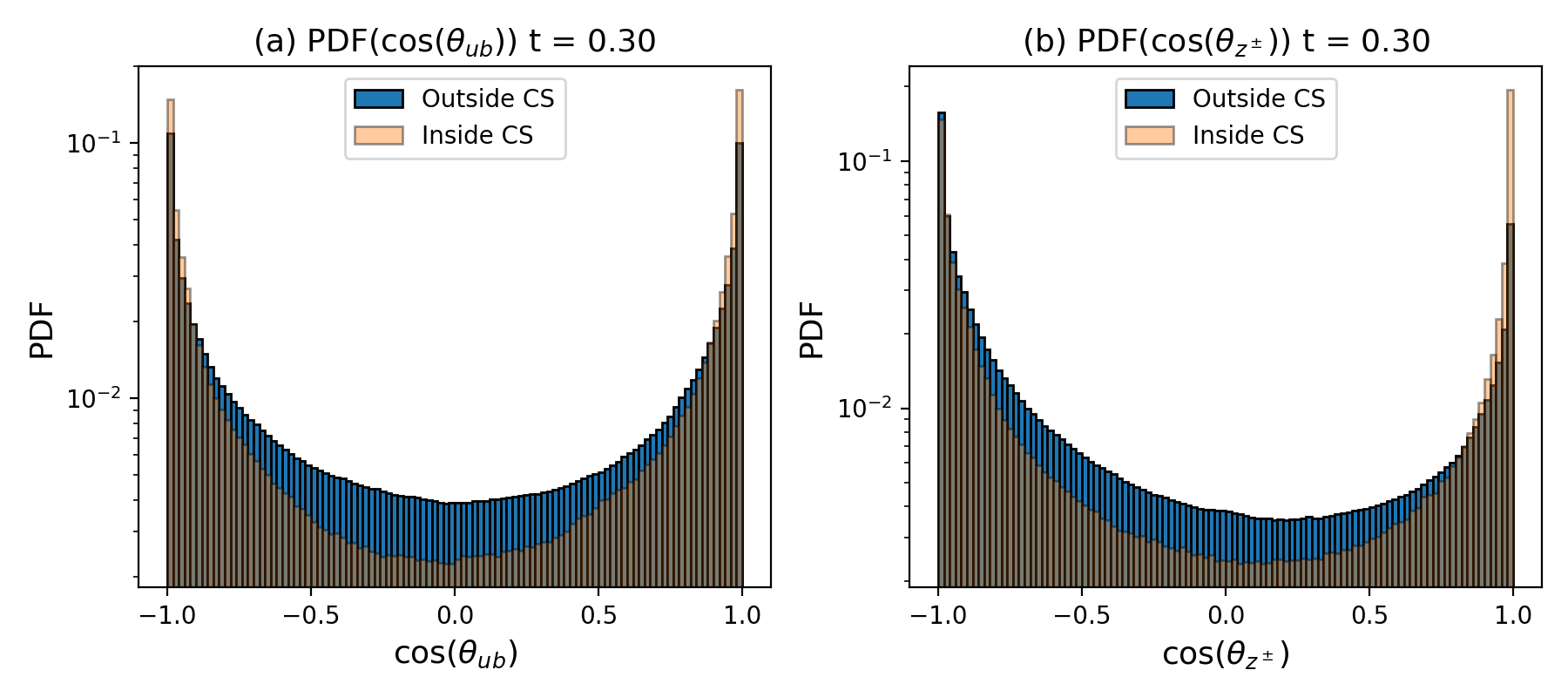}
    \caption{Probability distribution function (PDF) of $\cos(\theta_{ub})$ (a) and $\cos(\theta_{z^{\pm}})$ (b) at $t=0.3$. Here $\theta_{ub}$ is the angle between $\bm{u}$ and $\bm{b}$, and $\theta_{z^\pm}$ is the angle between $\bm{z^+}$ and $\bm{z^-}$. Background magnetic field was subtracted before calculating the angles. In each panel, blue bars correspond to regions outside current sheets and orange bars correspond to regions inside current sheets.
    }
    \label{fig:cos_ub_zpzm_distribution}
\end{figure*}


\begin{figure*}[htb!]
    \centering
    \includegraphics[width=\hsize]{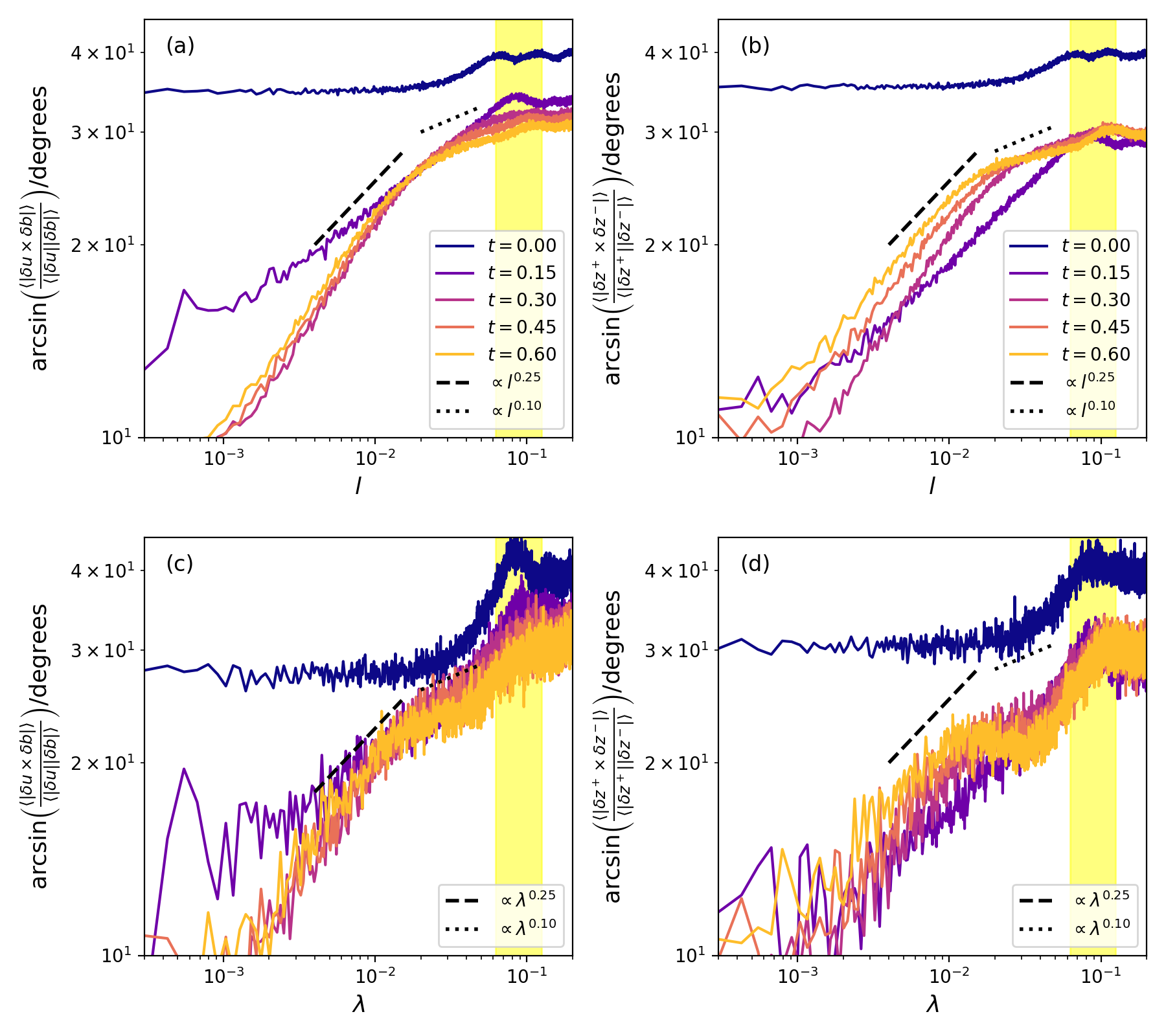}
    \caption{Scale-dependent alignment angles between (a) $\bm{u}$ and $\bm{b}$ and (b)  $\bm{z^+}$ and $\bm{z^-}$. Solid curves with different colors correspond to different times. Dashed and dotted lines show $\propto l^{0.25}$ and $l^{0.1}$ for reference. (c) \& (d): Same as (a) \& (b) but the horizontal axis is $\lambda$ (perpendicular to local $\bm{\delta b}$) instead of $l$.}
    \label{fig:scale_dependent_alignment_angle}
\end{figure*}

\section{Discussion: Current sheets and scale-dependent dynamic alignment}\label{sec:discussion}
In previous phenomenological models, current sheets in MHD turbulence are typically associated with the scale-dependent dynamic alignment.
In Figure~\ref{fig:cos_ub_zpzm_distribution}, we show the probability distribution functions (PDFs) of $\cos(\theta_{ub})$ and $\cos(\theta_{z^\pm})$ at $t=0.3$ over the whole simulation domain, where $\theta_{ub}$ and $\theta_{z^\pm}$ are the angle between $\bm{u}$ and $\bm{b}$ and angle between $\bm{z^+}$ and $\bm{z^-}$ respectively.
Blue bars and orange bars are calculated for regions outside and inside current sheets.
Overall, the two angles are concentrated at $0$ and $\pi$, both inside and outside current sheets, i.e. there is a trend that $\bm{u}$ and $\bm{b}$ are aligned with each other and so for $\bm{z^\pm}$.
Inside the current sheets these alignments are more evident than outside.
Moreover, panel~(b) shows that PDF($\cos(\theta_{z^\pm})$) outside current sheets is asymmetric, corresponding to the negative residual energy (Figure~\ref{fig:evolution_global_quantities}).
In contrast, the PDF inside current sheets is relatively more symmetric, indicating that current sheets are not a major contributor to the negative residual energy.
This result is consistent with 3D MHD simulations \citep{shi2025evolution} which show that there is no direct correspondence between the regions of negative residual energy and regions of strong intermittency.
We note that, although the alignment is stronger inside the current sheets, one cannot assert that there is a direct relation between the current sheets and the SDDA. 
In fact, in a reconnecting 2D current sheet, $\bm{u}$ and $\bm{b}$ are nearly-parallel with each other in most regions, thereby a current sheet naturally shows strong alignment between $\bm{u}$ and $\bm{b}$ and also between $\bm{z^+}$ and $\bm{z^-}$.

In Figure~\ref{fig:scale_dependent_alignment_angle}, we show the scale-dependent alignment angles. 
Panels~(a) and (b) show $\theta_{ub}(l)$ and $\theta_{z^\pm}(l)$, and panels~(c) and (d) show $\theta_{ub}(\lambda)$ and $\theta_{z^\pm}(\lambda)$.
Here $l=|\bm{l}|$ is non-directional spatial increment, and $\lambda$ is the spatial increment perpendicular to $\bm{\delta b}$, that is, the angle $\phi$ between $\bm{l}$ and $\delta \bm{b}(\bm{l})$ falls into the range $88^\circ-90^\circ$.
The alignment angles are calculated using
\begin{equation}
    \theta_{ub} = \arcsin \left( \frac{\langle \left| \delta \bm{u} \times \delta \bm{b} \right|\rangle}{ \langle \left| \delta \bm{u} \right| \left|  \delta \bm{b} \right| \rangle} \right)
\end{equation}
and similar for $\theta_{z^\pm}$.
We employ a Monte-Carlo method to evaluate the angles. That is, we randomly select two grid points $\bm{x_1}$ and $\bm{x_2}$ such that $\bm{l} = \bm{x_2} - \bm{x_1}$ and define $\delta \bm{b} = \bm{b}(\bm{x_2}) - \bm{b}(\bm{x_1})$ and same for $\delta\bm{u}$ and $\delta \bm{z^\pm}$.
We repeat this random process for $10^8$ times and calculate quantities like $\langle \left| \delta \bm{u} \times \delta \bm{b} \right| \rangle$, $\langle |\delta \bm{u}||\delta \bm{b}| \rangle$, and subsequently the alignment angles for the binned $l$ and $\phi$.
Figure~\ref{fig:scale_dependent_alignment_angle} confirm that both the alignment angles are smaller as we go toward smaller scales.
In the classic SDDA models \citep{boldyrev2006spectrum,loureiro2017role}, the alignment angle scales as $\theta_{ub} \propto \lambda^{0.25}$, while in the later model by \citet{chandran2015intermittency} the angles follow $\theta_{ub} \propto \lambda^{0.21}$ and $\theta_{z^\pm}\propto \lambda^{0.1}$.
Although these previous results are not directly applicable to the 2D turbulence, we still plot two reference lines with slopes of 0.25 and 0.10 respectively in each panel of Figure~\ref{fig:scale_dependent_alignment_angle} for visual assistance.
Panels~(a) and (b) show that, at early stage $t=0.15$, a single power law is observed for both the angles, and the slope for $\theta_{z^\pm}$ is steeper than that for $\theta_{ub}$.
At later times, double power-law relations form for both angles with a break scale at $l \sim (1-2) \times10^{-2}$.  
The large scale shows a slope close to $0.1$ and the small scale shows a slope close to $0.25$. 
Panels~(c) and (d) also show a clear break at $\lambda \sim (1-2) \times 10^{-2}$. The power-law slopes for small scales are similar to the top panels, but the slopes for large scales are quite different from the top panels.
What controls this break scale and how the alignment angles change with scale is still unclear. 
In the 2D compressible MHD simulations conducted by \citet{dong2018role}, such a break was not observed. One potential factor is the length of the current sheets, since Figure~\ref{fig:statistics_current_sheet}(c1-c3) show that the lower limit of the current sheet length is roughly $1\times10^{-2}$.
However, this is simply a hypothesis and further investigation is needed to fully understand this phenomenon.
Another noteworthy point is that, for $\theta_{z^\pm}(l)$, the alignment angle first evolves to a low level at $t=0.15$ and then gradually rises.
$\theta_{z^\pm}(\lambda)$ shows a similar evolution trend, though only for small scales.
This is reminiscent of the evolution of Kurtosis($\bm{b}$) shown in Figure~\ref{fig:kurtosis_b} and the residual energy shown in Figure~\ref{fig:evolution_global_quantities} and it implies that $\theta_{z^\pm}$ is likely affected by the evolution of turbulence current sheets.

\begin{figure}
    \centering
    \includegraphics[width=\linewidth]{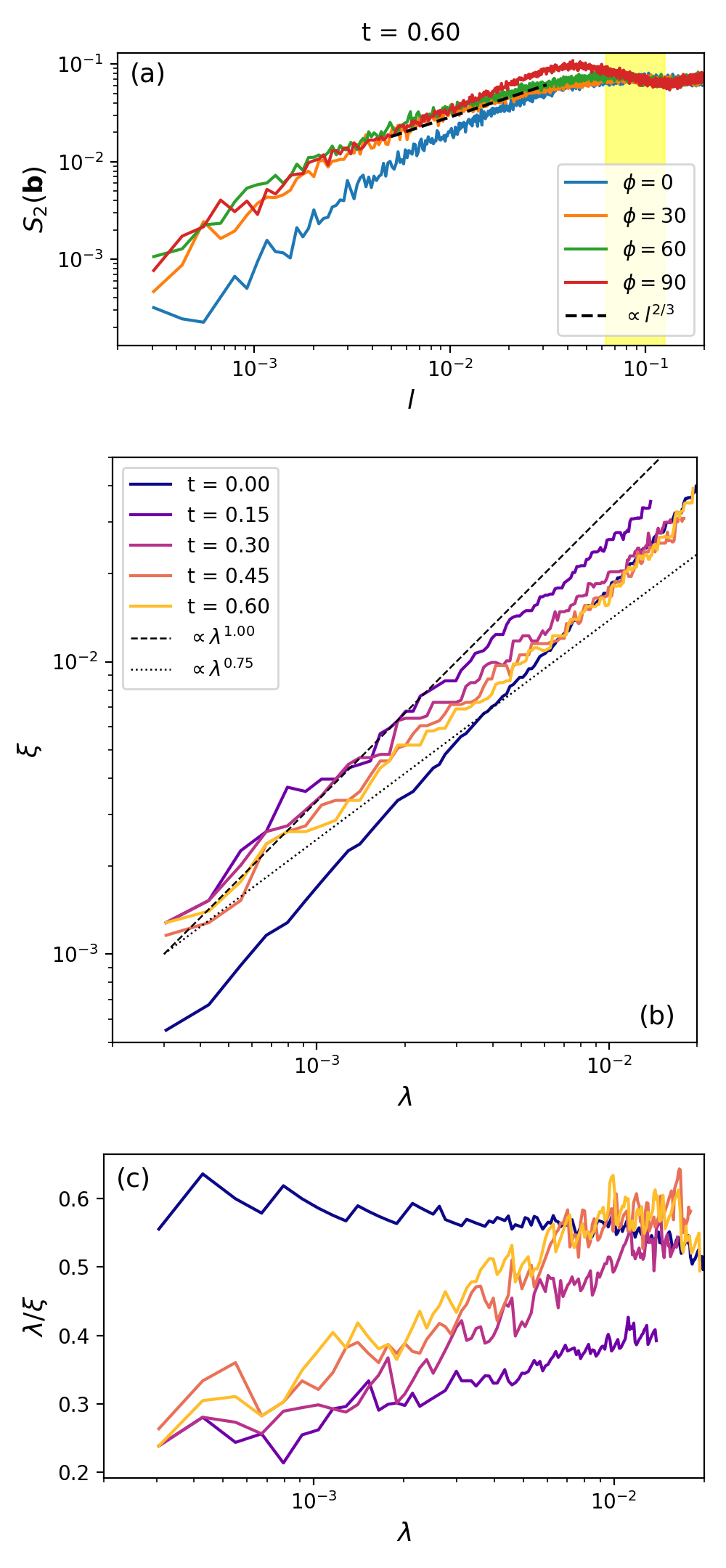}
    \caption{(a) Second-order structure functions of magnetic field $S_2(\bm{b})$ as a function of spatial increment $l$ at $t=0.6$. Different curves correspond to different angles between $\delta\bm{b}(\bm{l})$ and $\bm{l}$. (b) Scale parallel ($\xi$) to $\delta \bm{b}$ as a function of scale perpendicular ($\lambda$) to $\delta \bm{b}$. (c) $\lambda/\xi$ as a function of $\lambda$. }
    \label{fig:S2_lambda_xi}
\end{figure}

Last, we quantify the anisotropy of the turbulence ``eddies'' using the method described in \citep{sioulas2024higher}.
We apply the Monte-Carlo method to evaluate the second-order structure function of magnetic field while taking $\phi$ into consideration.
As shown by Figure~\ref{fig:S2_lambda_xi}(a), $S_2(\bm{b})$ is clearly anisotropic between the directions parallel ($\xi$, $\phi=0$) and perpendicular ($\lambda$, $\phi=90^\circ$) to the local magnetic field increment.
We numerically determine the function $\xi(\lambda)$ by equating $S_2(\bm{b},\xi)$ and $S_2(\bm{b},\lambda)$,
and the result is shown in panel~(b), which displays $\xi$ versus $\lambda$, and panel~(c), which displays $\xi/\lambda$ versus $\lambda$.
In general, $\xi > \lambda $, even at $t=0$, meaning that the turbulence eddies are stretched along the polarization direction.
$\lambda/\xi$ is roughly scale-independent at $t=0$, with values 0.5-0.6, but becomes scale-dependent soon after the simulation starts.
Similar to the Kurtosis shown in Figure~\ref{fig:kurtosis_b} and $\theta_{z^\pm}$ shown in Figure~\ref{fig:scale_dependent_alignment_angle}, $\lambda/\xi$ drops to its lowest value at $t=0.15$ and then starts to rise, implying that it is more or less related to the formation of turbulence current sheets.
However, we emphasize that, a temporal correlation or concurrence does not ensure causal relation.
The fact that current sheets, dynamic alignment, and turbulence eddy anisotropy evolve in a similar pattern does not mean that the current sheets are generated as a result of the dynamic alignment and eddy anisotropy.
In fact, the estimated $\lambda/\xi$ (Figure~\ref{fig:S2_lambda_xi}) are much larger than the aspect ratio of the current sheets (Figure~\ref{fig:statistics_current_sheet}(a1-a3)), which is mostly less than 0.1.
Similarly, \citet{dong2018role} concluded that the dynamic alignment angle is not small enough to cause the plasmoid instability.
In reality, the filling factor of the current sheets is very low ($<0.1$, Figure~\ref{fig:Jz2D_filling_factor_etc}) while the alignment angles and eddy anisotropy are evaluated over the whole simulation domain, meaning that the current sheet properties should have negligible contribution in the evaluation of the alignment angles and eddy anisotropy. Thus, it might be improper to take the alignment angles or eddy anisotropy as a proxy of current sheet geometry in the turbulence.

\section{Summary}\label{sec:summary}
In this study, we conducted a high-resolution 2D simulation of balanced incompressible MHD turbulence. We comprehensively analyzed the properties of current sheets generated during the turbulence evolution. 
The major findings are summarized below:
\begin{enumerate}
    \item The current sheets form much earlier ($t=0.15$) than one eddy turnover time ($t\approx 0.7$). The initial current sheet lengths are mostly comparable to the energy injection scales, i.e. they are controlled by the largest eddies in the system. Similar observations were reported in previous turbulence simulations \citep[e.g.][]{franci2015high,papini2019can}.
    \item As the current sheets continue to thin, tearing instability onsets when the current sheet aspect ratio approaches $S^{-1/2}$, generating smaller-scale current sheets.
    Again, we emphasize that the observed Sweet-Parker scaling does not necessarily hold in simulations with much higher Lundquist numbers, where an ideal-tearing scaling may emerge.
    \item Scale-dependent dynamic alignment is observed as shown by the alignment angles and eddy anisotropy. However, the alignment is much weaker than what is needed to explain the generated thin current sheets. 
\end{enumerate}
Our results indicate that there is no direct correspondence between the current sheet properties and SDDA.
Therefore, it is necessary to revisit the reconnection-mediated turbulence model. Particularly, we need to find the proper way to relate the current sheet properties with the turbulence diagnostics.
Additionally, we point out that, the results shown in Figure~\ref{fig:S2_lambda_xi}(b)\&(c) surprisingly resemble those in a high-resolution 3D MHD simulation of a balanced turbulence conducted by \citet{dong2022reconnection} (see their Fig.~3), including a scaling relation between $\xi$ and $\lambda$ with a slope slightly smaller than one, and very similar values of $\lambda/\xi$.
This similarity suggests that the $\xi-\lambda$ relation may not be uniquely tied to fully 3D effects, as assumed in SDDA models, and could instead reflect processes already captured in 2D. 

\begin{acknowledgments}
CS is supported by NSF SHINE \#2548299 and NASA ECIP \#80NSSC23K1064. The numerical simulations were conducted on Derecho: HPE Cray EX System (https://doi.org/10.5065/qx9a-pg09) of Computational and Information Systems Laboratory (CISL), National Center for Atmospheric Research (NCAR) (allocation No. UCLA0063).
\end{acknowledgments}

\software{Matplotlib \citep{Hunter2007Matplotlib}, NumPy \citep{harris2020array}, Scikit-Learn \citep{scikit-learn}, \texttt{LAPS} \citep{shi2024laps}}





\bibliography{references}{}
\bibliographystyle{aasjournalv7}



\end{CJK*}
\end{document}